\documentclass{kluwer}    

\newcommand{\Hya}{$\xi\,$Hya}
\newcommand{\muHz}{$\,\mu$Hz}

\usepackage{epsfig}

\begin{document}
\begin{article}

\begin{opening}         

\title{Simulating stochastically excited oscillations}
\subtitle{The mode lifetime of \Hya} 

\author{Dennis \surname{Stello}$^{1,2}$}  
\author{Hans \surname{Kjeldsen}$^{2,3}$}  
\author{Timothy R. \surname{Bedding}$^{1}$}  
\author{Joris \surname{De Ridder}$^{4}$}  
\author{Conny \surname{Aerts}$^{4}$}  
\author{Fabien \surname{Carrier}$^{5}$}  
\author{S{\o}ren \surname{Frandsen}$^{2}$}  
\institute{$^1$School of Physics, University of Sydney, NSW 2006,
Australia\email{stello@physics.usyd.edu.au}\\
$^2$Department of Physics and Astronomy, University of Aarhus, 8000 Aarhus C,
Denmark\\
$^3$Teoretisk Astrofysik Center, Danmarks Grundforskningsfond, 8000 Aarhus C,
Denmark\\
$^4$Instituut voor Sterrenkunde, Katholieke Universiteit Leuven, 3001 Leuven, Belgium\\
$^5$Observatoire de Gen\`eve, 1290 Sauverny, Switzerland}

\runningauthor{Dennis Stello et al.}
\runningtitle{Simulating stochastic excited oscillations}


\begin{abstract}
The discovery of solar-like oscillations in the giant star \Hya~
(G7III) was reported by \inlinecite{Frandsen02}. 
Their frequency analysis was very limited due to alias problems
in the data set (caused by single-site observations). The extent to which the 
aliasing affected their analysis was
unclear due to the unknown damping time of the stellar oscillation modes.
In this paper we describe a simulator created to generate time series of 
stochastically excited oscillations, which takes as input an arbitrary 
window function and includes both white and non-white noise.
We also outline a new method to compare a large number of simulated 
time series with an observed time series to determine the 
damping time, amplitude, and limited information on the degree 
of the stochastically excited modes.
For \Hya~we find the most likely amplitude to be $\sim 2\,$m/s, in good 
agreement with theory \cite{HoudekGough02}, and the most 
likely damping time to be $\sim 2\,$days, which is much shorter than the 
theoretical value of 15--$20\,$days calculated by \inlinecite{HoudekGough02}. 
\end{abstract}


\end{opening}

\section{Introduction}

The recent detection of solar-like oscillations in the G7 giant star \Hya~\\
\cite{Stello02,Frandsen02} promises new interesting prospects for asteroseismology 
in this part of the Hertzsprung-Russell diagram.
The amplitude spectrum of \Hya, based on a one-month time series measured in velocity, 
showed a clear excess of power
within a broad frequency envelope, similar to that seen in the Sun and other 
solar-type stars \cite{BeddingKjeldsen03}. 
The envelope was centered at $\sim90$\muHz, as expected from 
scaling the acoustic cut-off frequency of the Sun 
\cite{Brown91}. The highest peak in the amplitude spectrum was $\sim1.9\,$m/s.

Following these observational results, \inlinecite{HoudekGough02} calculated the 
theoretical damping rate, $\eta$, and amplitude for \Hya. The amplitude was 
$\sim2\,$m/s and, based on $\eta$, we calculated the damping time (or mode 
lifetime) as $1/(2\pi\eta)\sim15$--$20\,$days.

The autocorrelation function 
of the 
amplitude spectrum of \Hya~revealed a characteristic frequency separation of 
$~6.8$\muHz, in good agreement with the expected large frequency separation
of the radial modes (i.e. $l=0$) \cite{Stello02,Frandsen02}.
However, using the frequencies extracted from the observed amplitude spectrum, 
\inlinecite{Stello02} showed that a unique 
solution for the large frequency separation could not be found, due to 
aliasing. It seemed most likely, however, that the correct value was indeed in the 
range 6.8--7.0\muHz.
The observed frequencies could be explained by 
purely radial modes, but the presence of non-radial modes could not be excluded, 
again due to aliasing \cite{Stello02,Frandsen02}.

Based on a small sample of simulated time series, \inlinecite{Stello02} showed 
that the significance of the observed frequencies, due
to the effect from aliasing, was strongly dependent on the 
mode damping time adopted for \Hya.
Hence, in order to quantify the alias problems, the damping time for 
\Hya~has to be known.
However, as pointed out by \inlinecite{Stello02}, the damping time of \Hya~could not
be measured directly from the observed amplitude spectrum (by fitting Lorentzian 
profiles) because the power spectrum was too crowded. 
Neither was it possible to use the CLEANed spectrum to measure the damping time 
directly, since the number of frequencies and their position in the spectrum 
are not known.
Therefore, extensive simulations would be needed to estimate the damping time. 
Furthermore, it should be noted that the preliminary simulations performed by 
\inlinecite{Stello02} indicated that the damping time could be significantly 
shorter than 
predicted by theory \cite{HoudekGough02}. 
It is therefore important to establish a more precise determination of the 
damping time from the observations than was done by \inlinecite{Stello02}. 

In this paper we describe a time-series simulator, outlining 
the theoretical background and the technique used to simulate 
stochastically excited oscillations.
We use this simulator to determine the damping time of \Hya~using a new 
method, which is based on comparing the overall structure of the observed 
amplitude spectrum with a large sample of simulations. The method also gives
the amplitude and some limited information about the mode degree of the 
oscillations.

The outline of the paper is as follows: Section \ref{simulator} 
introduces the fundamental ideas used for simulating the stellar signal from 
stochastic pulsations, and describes the parameters that the simulator needs as input.
Section \ref{simhya} describes how the simulated time series of \Hya~were constructed 
and outlines how we decided the input parameters.
Based on a few results of the simulated data, we introduce in 
Sect. \ref{measureablecharacteristics} the method used to 
determine the damping time of \Hya~and present an optimum fit 
to the observations. Finally, we discuss the method and our results in 
Sect. \ref{discussion}, which also includes the conclusions.

\section{Simulator}\label{simulator}

The simulator described in this paper uses the same fundamental 
ideas (also described by \inlinecite{ChangGough98}) as the light 
curve simulator developed for the MONS and 
Eddington missions \cite{KjeldsenBedding98,Ridder02,Ridder03}.

\subsection{Stochastic excitation model}\label{stochastic_model}

The stellar signal, $S$, as a function of time, $t$, is modeled by 
\begin{equation}
  S(t) = \sum_{\nu=\nu_{1}}^{\nu_{n}}s_{\nu}(t)\, ,
\end{equation}
\noindent
where each $s_{\nu}(t)$ is a continuously re-excited damped harmonic 
oscillator that represents a single oscillation mode. 

In general, a damped harmonic oscillator without re-excitation can be 
expressed as 
\begin{equation}
\label{damposc}
  s_{\nu}(t) = A \sin(2\pi\nu t + \phi) e^{-t/d} \, ,
\end{equation}
\noindent
where $A$, $\nu$, $\phi$, and $d$ are the amplitude, frequency, phase, and 
damping time.
Rather than assigning a constant amplitude  to each oscillator, as in 
Eq. \ref{damposc}, we instead simulate the 
re-excitation and damping as a `kicking' and damping of the amplitude $A$. 
The amplitude of each mode is kicked independently at a rate characterized 
by the small time step, $\Delta t_{\mathrm{kick}}$, between each kick.
The independence of the re-excitation is established by having different  
phases, chosen at random, for each
mode, so that the time for each kick would not be simultaneous 
for the different modes.
After $n$ kicks 
the amplitude, $A$, assigned to a mode is 
\begin{equation}
\label{ar}
 A_{n} =  e^{-\Delta t_{\mathrm{kick}}/d} A_{n-1} + \varepsilon_{n} \, ,
\end{equation}
\noindent
where $e^{-\Delta t_{\mathrm{kick}}/d}$ is the damping factor and
$\varepsilon_{n}$ is the $n$th re-excitation kick, taken at 
random from a Gaussian distribution with zero mean and standard deviation 
$\sigma_{\varepsilon}$. 
In order to vary both amplitude and phase in time, we generate the time series as 
the sum of sine and cosine terms (using 
$A\sin(x+\pi/4)=A/\sqrt{2}\,(\sin(x)+\cos(x))$), which are simulated independently. 
Hence the expression for $s_{\nu}(t)$ used in this simulator is
\begin{equation}
\label{harmosc}
  s_{\nu}(t) = A_{n,1} \sin(2\pi\nu t + \phi_{\nu,1}) 
             + A_{n,2} \cos(2\pi\nu t + \phi_{\nu,2})\, .
\end{equation}
\noindent

The autoregressive process shown in Eq. \ref{ar} is asymptotically stable 
up to second order (i.e. it does not die out or `explode' as $n\to\infty$), 
provided $|e^{-\Delta t_{\mathrm{kick}}/d}|< 1$, which is always true for physically
meaningful values (i.e. $\Delta t_{\mathrm{kick}}> 0$ and $d> 0$)
\cite{Priestley81}.
In the asymptotic limit, when the process has relaxed, the mean 
is $\langle A_{n}\rangle=0$, and the variance, $\sigma_{A_{n}}^2$, 
can be expressed in terms of $\sigma_{\varepsilon}$ as
\begin{equation}
\label{sigmarel}
 \sigma_{A_{n}}^2 \simeq \frac{\sigma_{\varepsilon}^2}
                              {1-e^{-2\Delta t_{\mathrm{kick}}/d}}\, .
\end{equation}
\noindent

Using the amplitude-scaled discrete version of Parseval's theorem 
(see \opencite{KjeldsenFrandsen92}) we have for each term in Eq. \ref{harmosc}
\begin{eqnarray}
\label{parseval}
  P(\nu) &=& 2 \frac{1}{N}\sum_{n=1}^{N}|s_{\nu}(t_{n})|^2 \nonumber \\
         &=& 2 \frac{1}{N}\sum_{n=1}^{N}|A_{n}|^2 
                                       |\sin(2\pi\nu t_{n}+ \phi_{\nu})|^2 \nonumber \\
     &\simeq&  \frac{1}{N}\sum_{n=1}^{N}|A_{n}|^2 \nonumber \\
         &=&   \sigma_{A_{n}}^2
\end{eqnarray}
\noindent
for a large number of measurements, $N$, where $P(\nu)$ is the power 
at frequency $\nu$ in the power spectrum.

Finally, combining Eq. \ref{parseval} with Eq. \ref{sigmarel} gives the value of 
$\sigma_{\varepsilon}$ required to simulate a continuously
re-excited damped harmonic oscillator, as shown in Eq. \ref{harmosc}, that 
has an average power 
$P(\nu)=A(\nu)^2$ in the power spectrum. 
The resulting expression for $\sigma_{\varepsilon}$ is
\begin{equation}
\label{sigmavarepsilon}
 \sigma_{\varepsilon} = A \sqrt{\Delta t_{\mathrm{kick}}/d} \, ,
\end{equation}
\noindent
where the first-order approximation of $e^{x}$ has been used (which assumes 
$\Delta t_{\mathrm{kick}} \ll d$). We obtained Eq. \ref{sigmavarepsilon} by 
dividing $\sigma_{\varepsilon}$ by 
$\sqrt{2}$ to produce the correct standard deviation of the resulting amplitude 
for the combined sine and cosine term (cf. Eq. \ref{harmosc}). 

We repeat the autoregressive process (Eq. \ref{ar}) for a number of steps 
that is significantly longer than the characteristic relaxation time, 
to let it stabilize before the 
actual simulations of the stellar signal begin. As a rule of thumb, the minimum
number of initial steps of the autoregressive process (before starting the actual 
simulations) should correspond to at least
twice the damping time.

The simulated signal for each harmonic oscillator described above, 
$s_{\nu}(t)$ (Eq. \ref{harmosc}), may be viewed as the sum of two 
vectors in the complex 
plane. The main vector with length (amplitude) $A$ is anchored at the origin, 
cycling around it at 
the frequency of the oscillation. The  
excitation vector, with variable length, is anchored at the tip of the 
main vector. 
Its variable direction relative to the main vector can be separated into the 
two orthogonal components: phase (orthogonal to the main vector) and amplitude 
(parallel to the main vector).
\begin{figure}[t]\centering
\epsfxsize=12.0cm
\epsffile{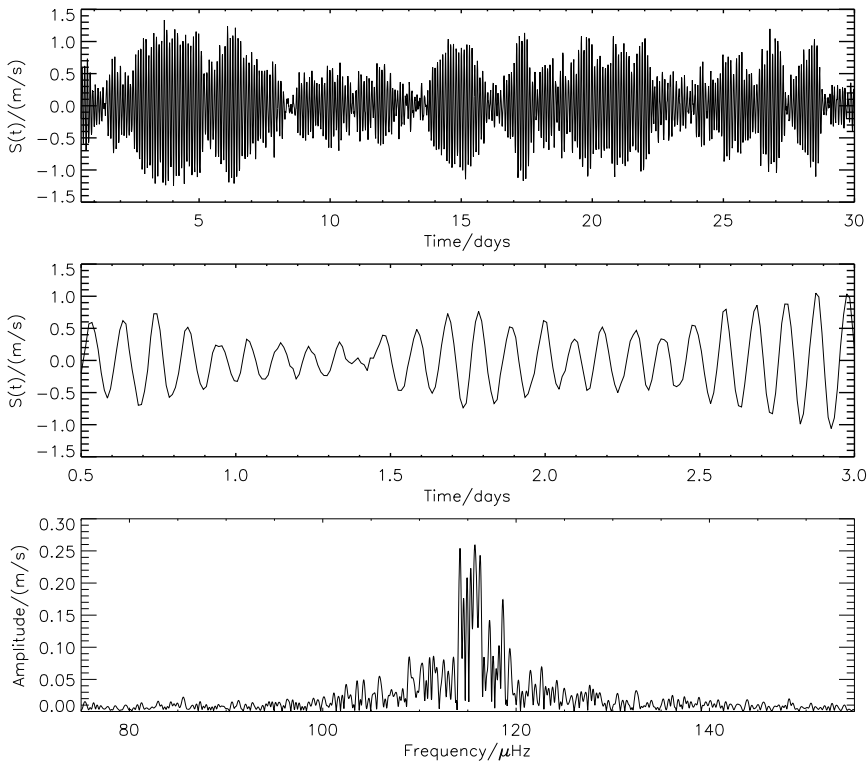}
\caption[]{\label{simexample}
\footnotesize{Simulated noise-free time series of a stochastically excited 
              and continuously damped oscillation at a single frequency
              (Eq. \ref{harmosc}).
              The input parameters are: $d=1.6\,$days, $\nu=115.7$\muHz, 
              $A=0.71\,$m/s, and $\Delta t_{\mathrm{kick}}=2\,$min.
              Top panel: Time series of 3000 data points sampled at
              $14.4\,$min. Middle panel: Close-up of the upper panel.
              Bottom panel: Amplitude spectrum of the time series.}}
\end{figure}
An example of such a stochastically excited oscillation at a single 
frequency is shown in Fig. 1. 
These simulations looks similar to observations of the Sun \cite{Leifsen01},
which supports the theory of continuous excitation by 
convective elements \cite{GoldreichKeeley77,Houdek99}.

\subsection{Amplitude interpolation}\label{parameterinterpolation}
The algorithm described in Sect. \ref{stochastic_model} calculates the 
amplitude (Eq. \ref{ar})
at regular time steps, but these do not necessarily coincide exactly with the 
times of the observations. 
Since the amplitude varies slowly, we simply use the value 
closest in time to each actual observation.
The sinusoidal oscillation itself (Eq. \ref{harmosc}), on the other hand, 
is evaluated at the exact times of the observations.
The result is a fast and reliable method for simulating stochastic excited
oscillations with an arbitrary observational window function.

\subsection{Noise calculation}
To make realistic simulations of stochastically excited oscillations, it is 
very important to be able to include noise \cite{Kjeldsen03}.
In our simulator both white noise and non-white noise are included.

The white noise is generated by a Gaussian random-number generator. 
For each data point, the
white noise is divided by the weight factor associated with that data point 
before it is added to the oscillation signal.

The non-white noise is created by first calculating the Fourier spectrum of 
a white noise source generated by a random generator. Then we multiply it by a 
function that describes the desired profile of the non-white noise. 
The result is then converted back to the time domain,  producing
the time series of the non-white noise source, which finally is added 
to the oscillation signal.

\subsection{Input parameters}
To summarize, the following input parameters have to be supplied to the 
simulator:
\begin{enumerate}
 \item A set of frequencies. 
 \item The amplitude corresponding to each frequency.
 \item The damping time corresponding to each frequency.
 \item The time of each observation (the observational window).
 \item The weight of each observation.
 \item A set of parameters defining the noise levels according to the chosen
       noise function.
 \item The time step between each re-excitation kick $\Delta t_{\mathrm{kick}}$.
\end{enumerate}

\section{Simulating time series of \Hya}\label{simhya}

We used the simulator described in Sect. \ref{simulator} to generate a large 
number of time series similar to that obtained for \Hya~by \inlinecite{Frandsen02}.
The input parameters for the simulator were as follows:

\begin{enumerate}
 \item Since the observations can be explained as purely radial modes 
   \cite{Stello02,Frandsen02}, this simple case 
   has been chosen for the main part of the current investigation. The input 
   frequencies, $\nu_{1},\ldots,\nu_{n}$, were the radial modes from a pulsation
   model \cite{Stello02}, rescaled so that the mean frequency separation was 
   6.8\muHz, in agreement with the observations \cite{Stello02,Frandsen02}. 

 \item    The relative amplitude of each 
   mode was defined according to an envelope (see Fig. \ref{inputampfreq}, 
   solid vertical lines).
   The envelope was obtained by smoothing the observed amplitude spectrum, 
   subtracting the 
   noise background and normalizing the peak to unity.
   In order to match simulations with observations the width 
   of the envelope was made adjustable. We chose to do this by raising the curve 
   to the $x$th power, 
   where $x$ was a free parameter, which was a convenient way of changing
   the width of the envelope without changing the height. 
   The envelope was then scaled vertically by a factor 
   $\mathrm{Amp}_{\mathrm{scale}}$, which was a free parameter that 
   represents the oscillation amplitude of \Hya. 
   Figure. \ref{inputampfreq} (dashed curve) shows the normalized envelope for $x=2$
   and $\mathrm{Amp}_{\mathrm{scale}}=1\,$m/s.
   \begin{figure}[h]\centering
   \epsffile{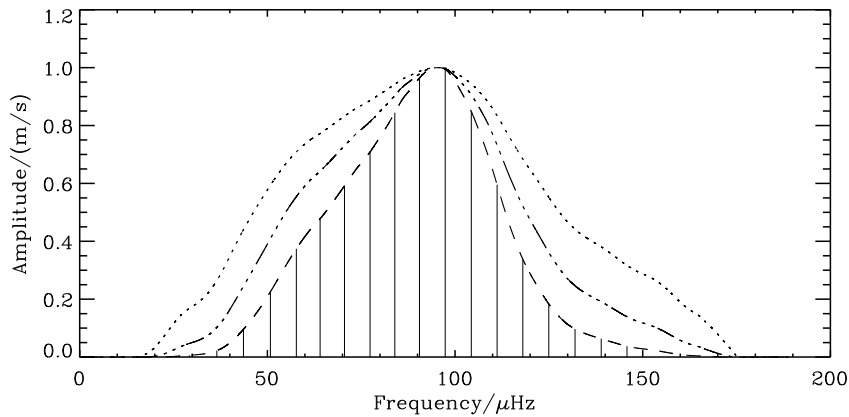}
   \caption[]{\label{inputampfreq}
   \footnotesize{Three amplitude envelopes: dashed curve: $x=2$ 
                 ($\mathrm{FWHM_{\mathrm{in}}}=48.0$\muHz); 
                 dashed-dot-dot-dot curve: $x=1.2$ 
                 ($\mathrm{FWHM_{\mathrm{in}}}=64.0$\muHz); 
                 dotted curve: $x=0.7$ 
                 ($\mathrm{FWHM_{\mathrm{in}}}=81.6$\muHz). The 
                 vertical solid lines indicate the mode frequencies.}}
   \end{figure}

 \item The damping time, $d$, was a free 
   parameter and was the same for all modes. This is probably a reasonable 
   approximation, since the theoretical damping rate shows a 
   flat plateau covering a fairly broad range of frequencies ($\sim70$--130\muHz) 
   \cite{HoudekGough02}.  

 \item The observational window function was exactly the same as for the 
   actual observations \cite{Stello02,Frandsen02}. The amplitude spectrum of 
   the observational window 
   (the spectral window) is shown in Fig. 3. 

   \begin{figure}[t]\centering
   \epsfxsize=12.0cm
   \epsffile{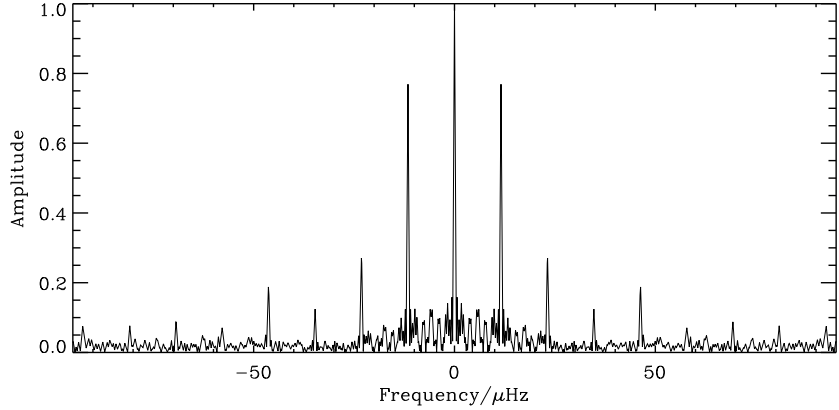}
   \caption[]{\label{windowfunction}
   \footnotesize{Spectral window of the \Hya~time series.}}
   \end{figure}

 \item The weight of each observation was $1/\sigma_{i}^2$, where $\sigma_{i}$ 
   was the noise associated with each observed data point.

 \item The noise was the sum of a white and a non-white noise component, where  
   the latter was described by a linear model. The noise was therefore specified by 
   three parameters: the white noise level ($\mathrm{Noise}_{\mathrm{white}}$), 
   the slope of the non-white noise
   ($\mathrm{Noise}_{\mathrm{slope}}$), and a scaling of the non-white noise
   ($\mathrm{Noise}_{\mathrm{scale}}$). 

 \item The time step $\Delta t_{\mathrm{kick}}$ was set to $2\,$min to meet 
   the requirement that $\Delta t_{\mathrm{kick}} \ll d$ for all damping times tested
   in this investigation.
\end{enumerate}

Before starting the actual simulations, we repeated the autoregressive process 
(Eq. \ref{ar}) for 120000 steps, 
corresponding to 167 days, which is significantly longer than the characteristic 
relaxation time for all the damping times tested in this investigation, 
to ensure it had stabilized.

\section{Method and results}\label{measureablecharacteristics}

\begin{figure}[h]\centering
\epsffile{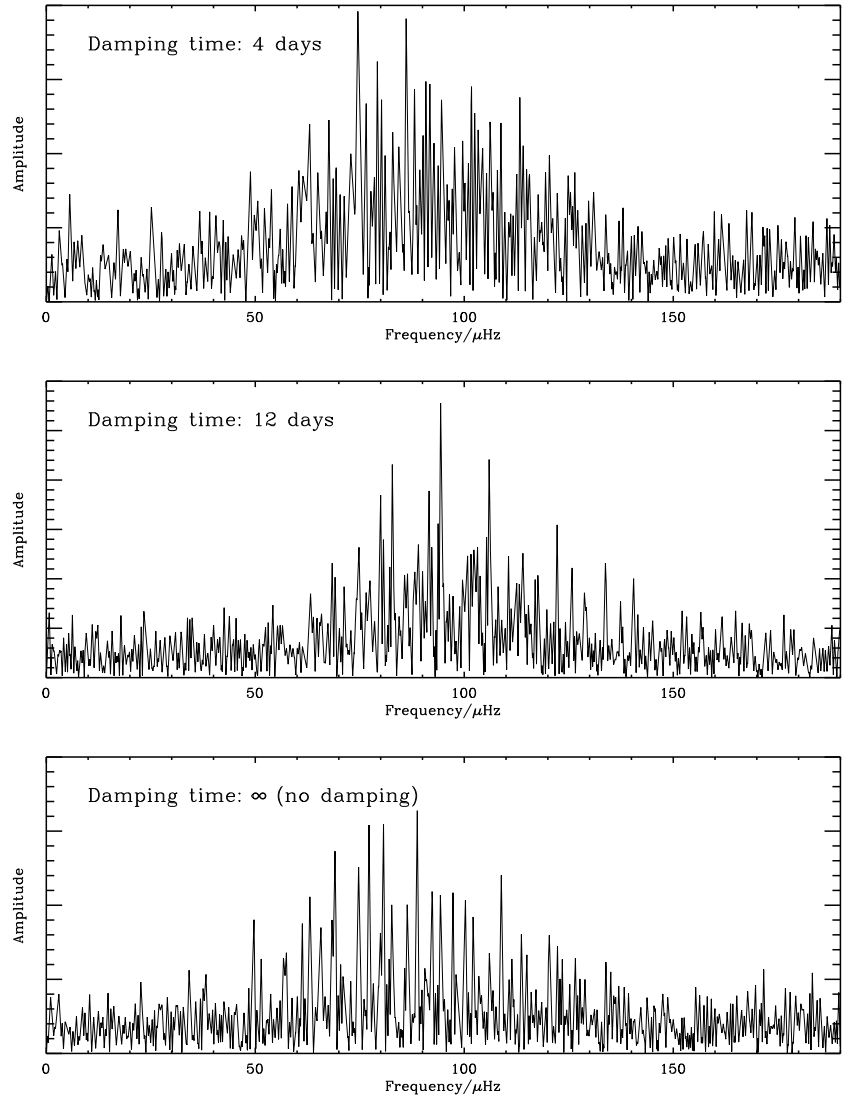}
\caption[]{\label{sim_amp_eks}
\footnotesize{Amplitude spectra based on simulated time series. The input damping 
              times for the time series are shown in each panel. For infinite 
              damping time (bottom panel)
              the input amplitudes were randomized.}}
\end{figure}

Examples of simulated amplitude spectra with different damping times are shown 
in Fig. \ref{sim_amp_eks}.
It is clear that amplitude spectra based on a short damping time 
have much more densely packed peaks (top panel) than do those with long damping times 
(bottom panel). 
This arises because the continuous re-excitation of modes introduces 
slight shifts in the phase as a function of time, which shows up in the amplitude 
spectrum as extra peaks slightly offset in frequency.
The fact that amplitude spectra based on different damping times 
display such different characteristics,  as 
seen in Fig. \ref{sim_amp_eks}, suggests the possibility of 
measuring the mode damping time from the overall 
structure of the amplitude spectrum.

We determined how well the simulated time series 
reproduced the observations by using eight measurable 
parameters that characterized different features of the amplitude spectra.
\begin{figure}[t]\centering
\epsffile{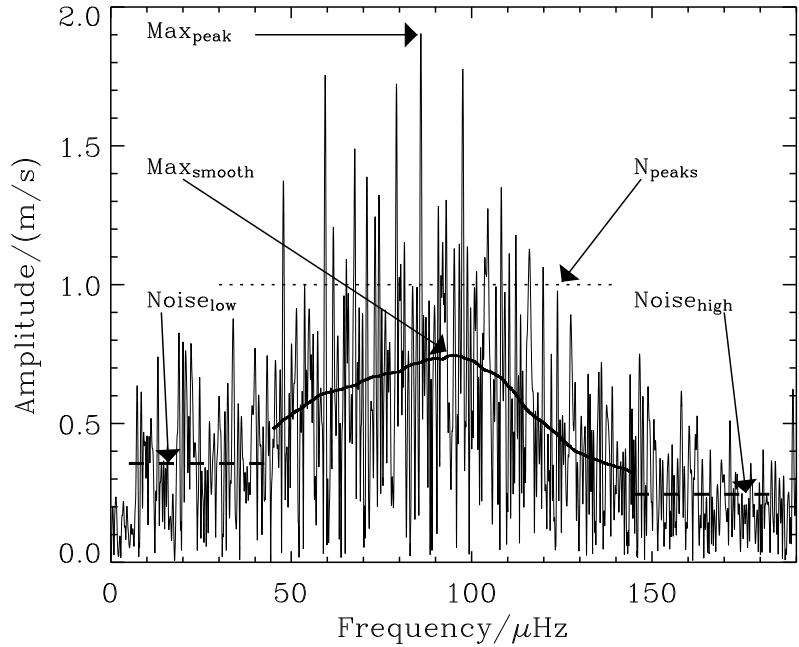}
\caption[]{\label{measureablecharacteristics}
\footnotesize{Amplitude spectrum (observed) of \Hya. Different 
              parameters characterizing the spectrum are indicated 
              (see text page \pageref{parameters}).}}
\end{figure}
In Fig. 5 
we show the observed amplitude spectrum of \Hya, 
together with the measurable parameters that specify 
different characteristics of the amplitude spectrum. 

\label{parameters} The highest peak ($\mathrm{Max}_{\mathrm{peak}}$) is the 
highest amplitude found 
in the spectrum, while $\mathrm{Max}_{\mathrm{smooth}}$ is the height  
of the smoothed amplitude spectrum, where smoothing was done twice with a boxcar filter.
The widths of the boxcars were 20\muHz~followed by 5\muHz.  
The two noise levels, $\mathrm{Noise}_{\mathrm{low}}$ and 
$\mathrm{Noise}_{\mathrm{high}}$ are measured as the mean amplitude in the 
frequency ranges 5--45\muHz~and 145--185\muHz, respectively.
The number of detected peaks with amplitudes above a threshold of 
$1.0\,$m/s (S/N$\ge 3.5$),  
denoted N$_{\mathrm{peaks}}$, is found by CLEANing the amplitude spectrum until 
the amplitude threshold has been reached. 
Our CLEAN process subtracts one frequency at a time (the one with the highest 
amplitude), but recalculates the amplitude, phase and frequencies of the previously 
subtracted peaks while fixing the frequency of the latest extracted peak.
In this way, the fit of sinusoids to the time series is done 
simultaneously for all peaks.
The mean amplitude is measured both over the entire frequency 
range 0--190\muHz~(Amp$_{\mathrm{tot}}$) and in the central part that is
dominated by the stellar excess of power, 40--140\muHz~(Amp$_{\mathrm{cen}}$).
Finally, the width of the excess power (W$_{\mathrm{env}}$) is measured as the 
FWHM of the smoothed spectrum after the noise has been subtracted. Due to the window 
function, stellar excess power `leaks' into the frequency regions where 
we measure the noise. The subtraction of noise in the determination of the 
envelope width therefore includes some stellar power, making W$_{\mathrm{env}}$ 
smaller than the input widths (Fig. \ref{inputampfreq}), which 
therefore should not be compared.

We now describe how the eight parameters measured 
from the simulated amplitude spectra ($\mathrm{Max}_{\mathrm{peak}}$, 
$\mathrm{Max}_{\mathrm{smooth}}$, $\mathrm{Noise}_{\mathrm{low}}$, 
$\mathrm{Noise}_{\mathrm{high}}$, N$_{\mathrm{peaks}}$, Amp$_{\mathrm{cen}}$, 
Amp$_{\mathrm{tot}}$, W$_{\mathrm{env}}$) 
are affected by changes in the 
input parameters, namely damping time ($d$), noise 
($\mathrm{Noise}_{\mathrm{white}}$, 
$\mathrm{Noise}_{\mathrm{slope}}$, $\mathrm{Noise}_{\mathrm{scale}}$),
amplitude ($\mathrm{Amp}_{\mathrm{scale}}$),  
envelope width (determined by the 
exponent $x$), and input frequencies ($\nu_{1},\ldots,\nu_{n}$).

\begin{figure}[p]\centering
\epsffile{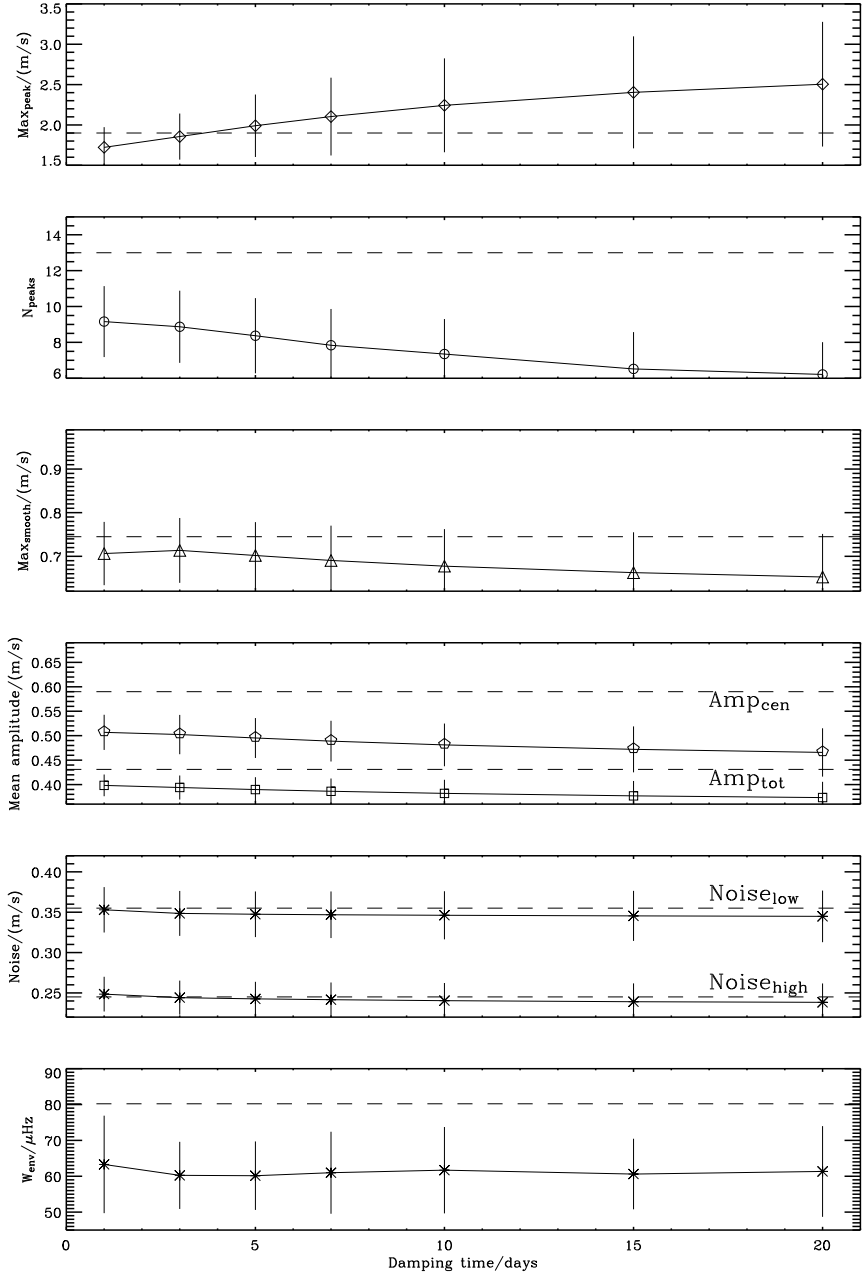}
\caption[]{\label{dampingtimesensitivity}
\footnotesize{The eight measured parameters
              as a function of the damping time. The data points are based on 
              7 distinct sets of simulated spectra only differing in their damping 
              time. Each set comprised 100 independent simulated time series. 
              Each plotted
              point is the mean of the values found from the 100 simulations 
              within each set, and the vertical bars indicate the rms scatter that
              results from the stochastic nature of the 
              simulation. The points are connected by solid lines to guide the eye. 
              The \textit{dashed lines} denote the 
              values measured from the observations. The common input parameters 
              for the 7 sets are: $\mathrm{Amp}_{\mathrm{scale}}=2.1\,$m/s, 
              $x=2.0$ (FWHM$_\mathrm{in}=48.0$\muHz), 
              $\mathrm{Noise}_{\mathrm{white}}=0.20\,$m/s,
              $\mathrm{Noise}_{\mathrm{slope}}=1.5$,
              $\mathrm{Noise}_{\mathrm{scale}}=3.0$, and 18 input frequencies
              with a mean frequency 
              separation of 6.8\muHz.}}
\end{figure}

\subsection{Damping time}
Figure 6 
shows the dependence of each of the eight 
measured parameters on the damping time. Values adopted for the other input
parameters are given in the figure caption.
Each plotted point is the mean of 100 simulations 
with the same input parameters but different random number seeds, and the vertical
bars show the rms scatter over these 100 simulations. 
This rms is the intrinsic scatter  
due to the stochastic nature of the oscillations, and is therefore the quantity 
we should use to decide whether the simulations match the observations.
The parameters measured from the observed amplitude spectrum are 
indicated by horizontal dashed lines.

All the parameters increase as the damping time gets shorter except 
for $\mathrm{Max}_{\mathrm{peak}}$, which falls off. The fall in 
$\mathrm{Max}_{\mathrm{peak}}$ is simply because the re-excitation spreads
the power over more peaks, giving less power in each.
The relative change in most of the parameters over the range of damping times 
plotted in Fig. 6 
is fairly small 
($\lesssim10\%$) and is generally less than the rms scatter. These 
parameters are therefore not very sensitive measures of the damping time but should
still be matched with the observed values to constrain the other input parameters. 
However, $\mathrm{Max}_{\mathrm{peak}}$ and N$_{\mathrm{peaks}}$ change by roughly
50\% in the same range and in opposite directions, making them the obvious  
parameters of choice for constraining the 
mode damping time. 
The correlation coefficient between $\mathrm{Max}_{\mathrm{peak}}$ and 
N$_{\mathrm{peaks}}$ is only
$\rho\sim0.10$--0.15, based on a few sets of 100 simulated amplitude spectra, where 
each set had different input parameters.
These two parameters can therefore be regarded as uncorrelated.

The 2nd, 3rd, and 4th panels in Fig. 6 
all indicate that, 
for all damping times, the amount of power in these simulations is too low.
The bottom panel shows that this is partly because the width is too narrow. 
The input power can be adjusted by changing the input noise, amplitude, 
envelope width, and the number of frequencies (their mean separation). 
We address each of these possibilities in turn in the next sections.

\subsection{Noise}
We investigated how the measured parameters changed as a function of 
input noise. We only show the results of changes in $\mathrm{Noise}_{\mathrm{white}}$
because this parameter gave us all the control we needed to adjust the simulations to match the noise level in the observations.
The two other noise parameters were fixed at the values used for 
Fig. 6. 

\begin{figure}[p]\centering
\epsffile{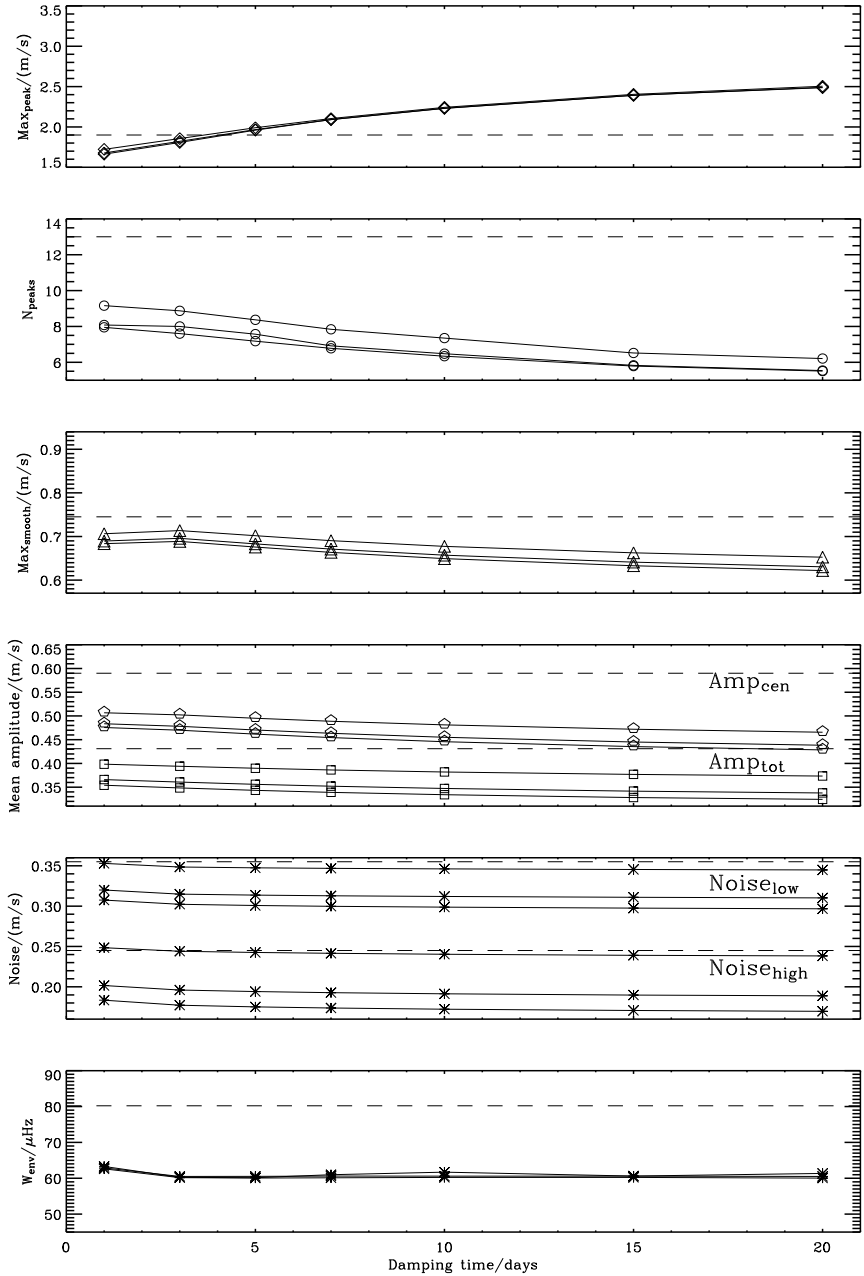}
\caption[]{\label{noisesensitivity}
\footnotesize{The effect of changing the white noise. 
              These panels are similar to those plotted in Fig. 6, 
              but for  
              three different values of the white noise.
              Each measured parameter 
              is therefore shown as three points for every damping time.
              The values for input noise (from top to bottom in each panel) are: 
              $\mathrm{Noise}_{\mathrm{white}}=0.2\,$m/s 
              (as in Fig. 6), 
              $\mathrm{Noise}_{\mathrm{white}}=0.1\,$m/s, and
              $\mathrm{Noise}_{\mathrm{white}}=0.0\,$m/s. All parameters
              show an decrease as $\mathrm{Noise}_{\mathrm{white}}$ decreases.
              Note that the y-axis has been shifted, but not scaled, relative to 
              Fig. 6. 
              For clarity, the vertical bars indicating rms scatter have been omitted.
              }}
\end{figure}
Figure 7 
shows that the parameters most sensitive to changes 
in the input noise are 
$\mathrm{Noise}_{\mathrm{low}}$ and $\mathrm{Noise}_{\mathrm{high}}$, while the other
parameters show much smaller changes. 
The input noise can therefore be regarded as the final fine tuning parameter, and has 
less importance for constraining the other input parameters. The panel showing 
the measured noise will therefore be omitted in the following plots.

\subsection{Amplitude}\label{amplitude}
\begin{figure}[p]\centering
\epsffile{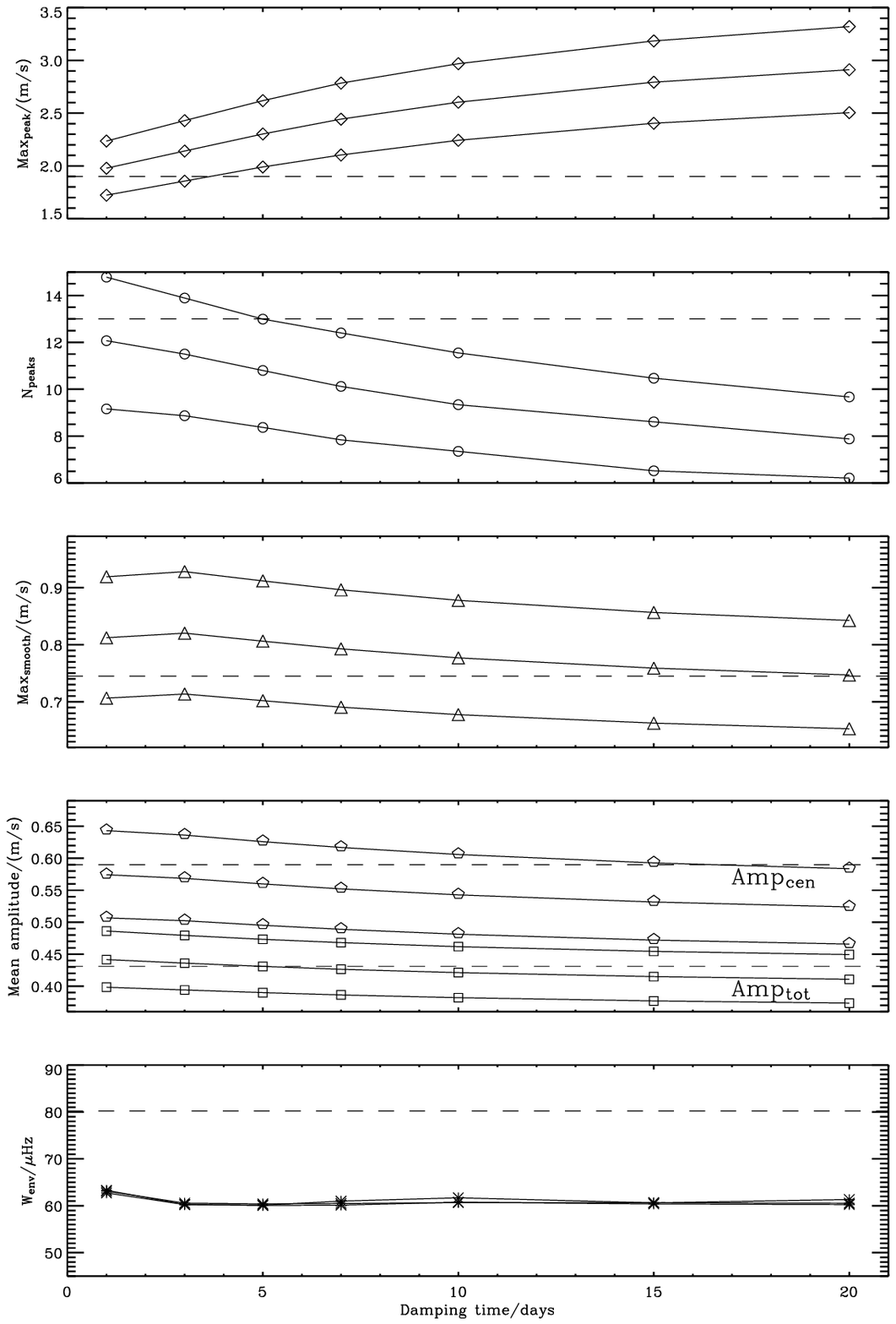}
\caption[]{\label{amplitudesensitivity}
\footnotesize{The effect of changing the input amplitude.
              These panels are similar to those plotted in Fig. 6, 
              but for three different values of input amplitude. 
              Each measured parameter 
              is therefore shown as three points for every damping time.
              The input amplitudes (from bottom to top in each panel) are: 
              $\mathrm{Amp}_{\mathrm{scale}}=2.1\,$m/s 
              (as in Fig. 6), 
              $\mathrm{Amp}_{\mathrm{scale}}=2.5\,$m/s, and
              $\mathrm{Amp}_{\mathrm{scale}}=2.8\,$m/s. All parameters
              show an increase as $\mathrm{Amp}_{\mathrm{scale}}$ increases.
              }}
\end{figure}
In Fig. 8 
we show the dependence of the measured parameters 
on the input amplitude. It can be seen that all parameters show an 
increase with input amplitudes, which is expected due to the increase in 
power, although the change is very small for W$_{\mathrm{env}}$. 
 
It is clear from  Fig. 8 
that the input amplitude is 
constrained by $\mathrm{Max}_{\mathrm{peak}}$, which should not be too high,
and by N$_{\mathrm{peaks}}$, which should not be too low. 
The difficulty in satisfying both constraints simultaneously becomes greater for 
larger values of the damping time.

\begin{figure}[p]\centering
\epsffile{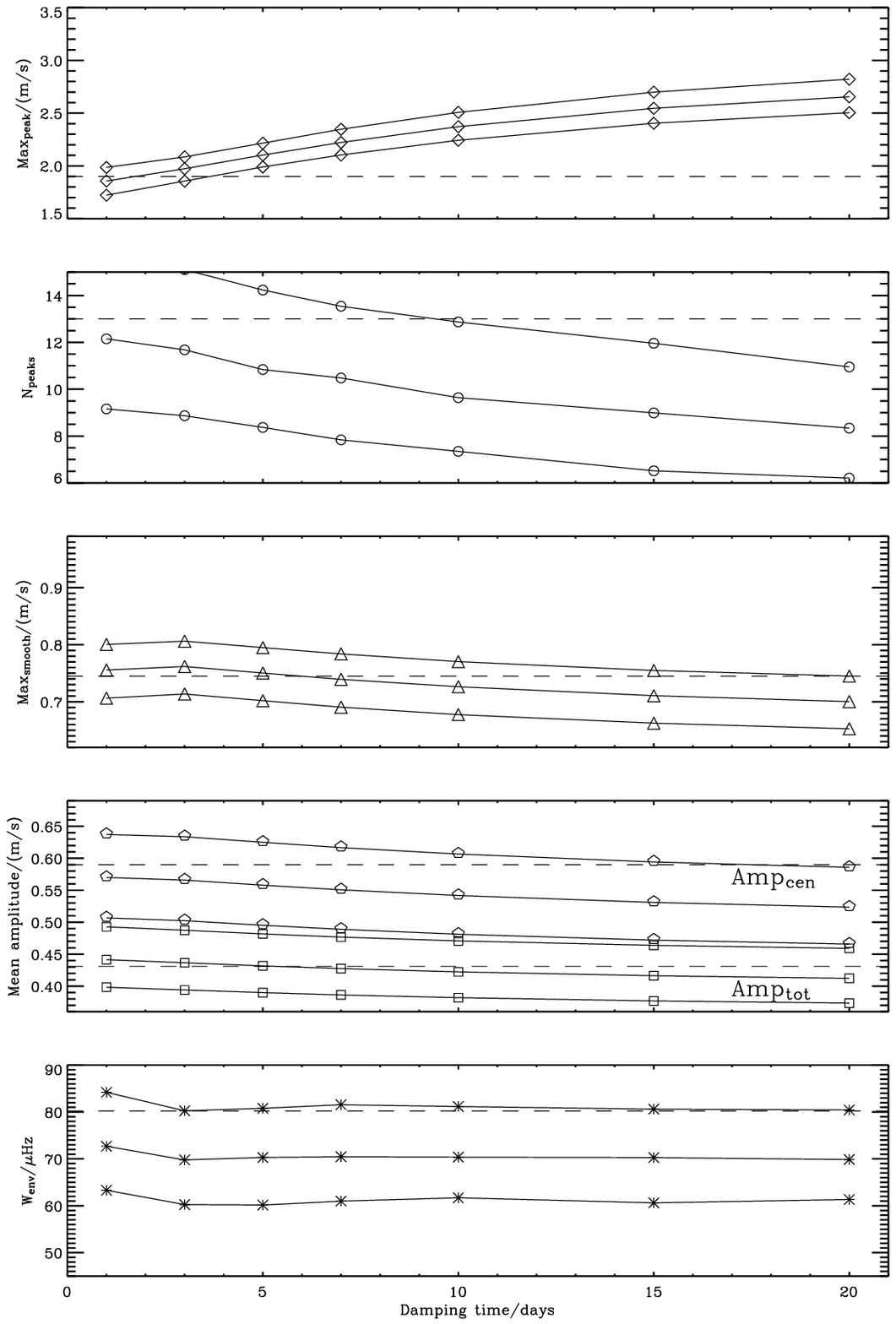}
\caption[]{\label{widthsensitivity}
\footnotesize{The effect of changing the envelope width. These panels are similar 
              to those plotted in Fig. 6, 
              but for three different values of envelope width. 
              Each measured parameter 
              is therefore shown as three points for every damping time.
              The input widths (from bottom to top in each panel) are: 
              $x=2$ (FWHM$_{\mathrm{in}}=48.0$\muHz) 
              (as in Fig. 6), 
              $x=1.2$ (FWHM$_{\mathrm{in}}=64.0$\muHz), and
              $x=0.7$ (FWHM$_{\mathrm{in}}=81.6$\muHz). All parameters
              show an increase as the envelope width increases. 
              }}
\end{figure}

\begin{figure}[p]\centering
\epsffile{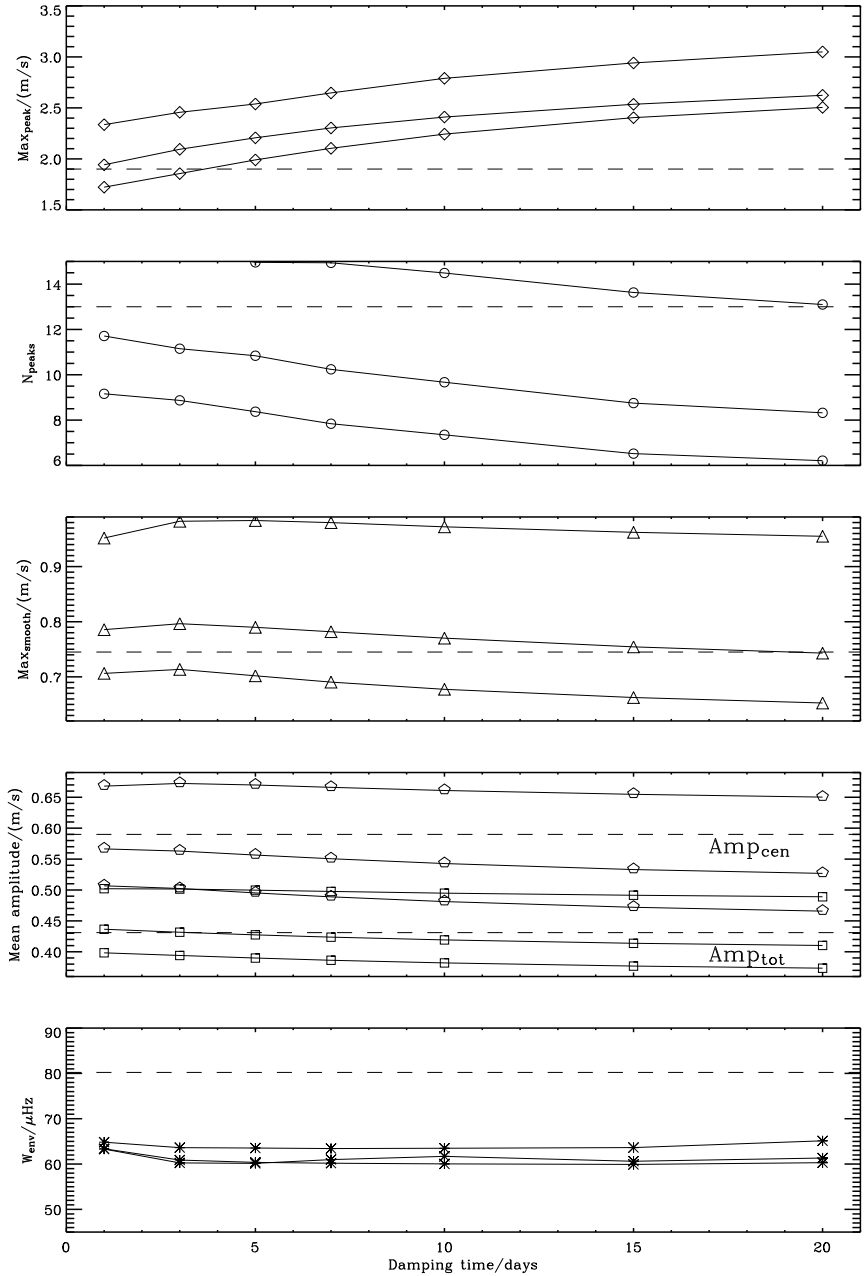}
\caption[]{\label{freqsepsensitivity}
\footnotesize{The effect of changing the number of input frequencies. 
              These panels are similar to those plotted in 
              Fig. 6, 
              but for three different values of the frequency separation.
              Each measured parameter 
              is therefore shown as three points for every damping time.
              The frequency separations (from bottom to top in each panel) are: 
              6.8\muHz~(as in Fig. 6), 
              5.0\muHz, and 3.4\muHz. All parameters
              show an increase as the frequency separation decreases. 
              The 3.4\muHz~simulations corresponds to having one non-radial
              mode per radial mode positioned halfway between each radial mode.
              }}
\end{figure}

\subsection{Envelope width}\label{envelopesensitivity}
Another way of injecting more power into the simulations, enabling more peaks to
be detected without having $\mathrm{Max}_{\mathrm{peak}}$ increase significantly,
is by making the envelope wider (cf. Fig. \ref{inputampfreq}).
Figure 9 
shows the dependence of the measured parameters 
on the envelope width. As expected, an increase in envelope width
gives rise to an increase in all parameters, due to the power increase. 
$\mathrm{Max}_{\mathrm{peak}}$ is, however, nearly unaffected.
Thus, by increasing the width we  construct simulations
from which more peaks are detected (N$_{\mathrm{peaks}}$ increases), without  
affecting $\mathrm{Max}_{\mathrm{peak}}$, hence making the 
simulations better fit the observations,
which is also supported by 
the bottom panel. Since the input width is the only input parameter to induce
significant changes in W$_{\mathrm{env}}$, we see from the bottom panel that 
the input width can be fixed at a value around 80\muHz.

\subsection{Input frequencies}\label{frequencies}

Finally, we tested the dependence of the measured parameters on the number of 
input frequencies. The above examples have included only the 
radial modes, so we have simulated the presence of non-radial modes by  
decreasing the mean frequency separation. 
Since we used the radial modes from a pulsation model \cite{Stello02}, they showed 
a small scatter of 0.2\muHz~from perfectly uniform spacing. This scatter was kept 
constant as we reduced the frequency separation and included more frequencies.

The results of the change in frequency separation are shown in 
Fig. 10. 
As expected we see an increase in all measured parameters when increasing 
the number of input frequencies (decreasing the frequency separation), due to the
increase in the power, although W$_{\mathrm{env}}$ is nearly unaffected.

\subsection{Finding the optimum input parameters}\label{dampingtime}

We find the most likely damping time by examining 
$\mathrm{Max}_{\mathrm{peak}}$/$\mathrm{Max}_{\mathrm{smooth}}$, 
which is relatively insensitive to $\mathrm{Amp}_{\mathrm{scale}}$, envelope width, 
input frequencies, and noise, but not to the damping time 
(see Fig. 11). 
\begin{figure}[p]\centering
\epsffile{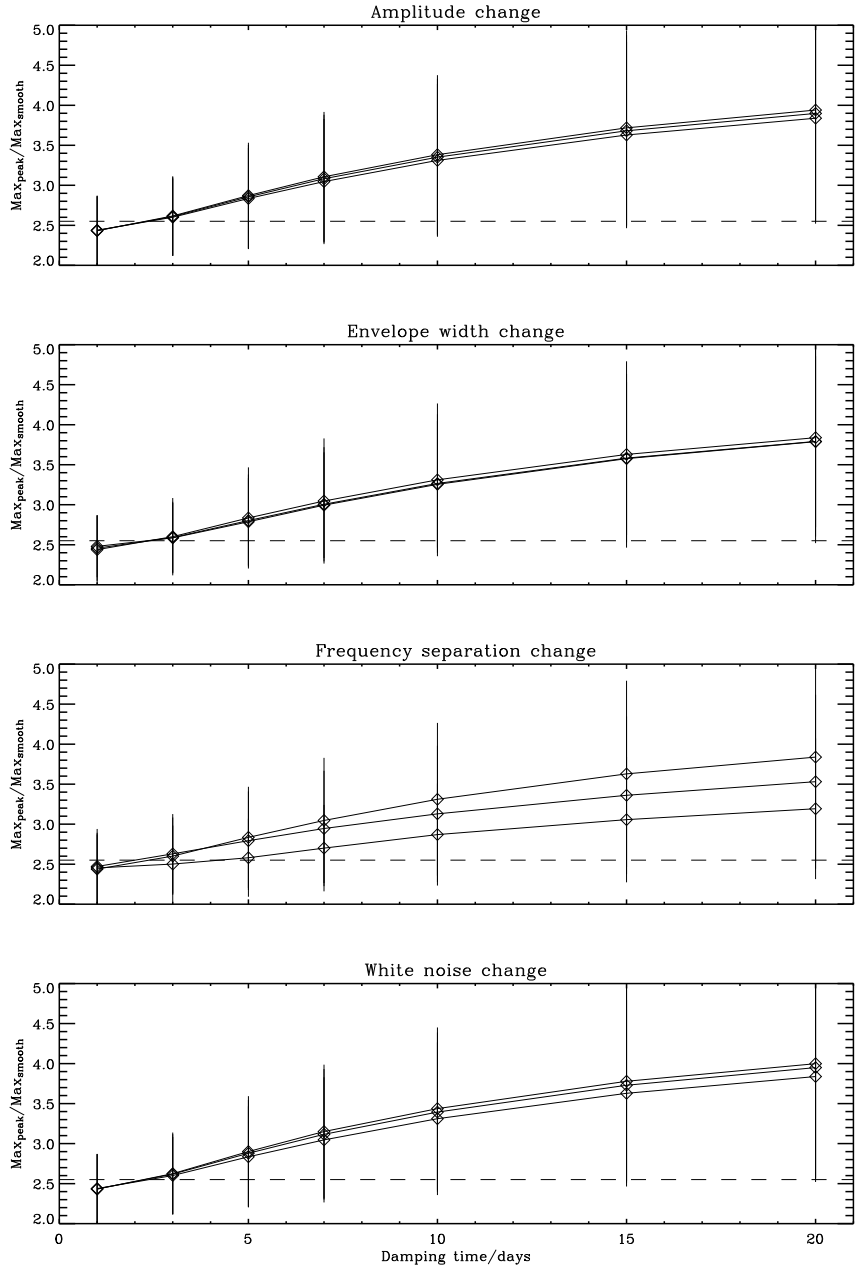}
\caption[]{\label{maxmedsensitivity}
\footnotesize{The ratio 
              $\mathrm{Max}_{\mathrm{peak}}$/$\mathrm{Max}_{\mathrm{smooth}}$
              as a function of damping time. The four panels show the sensitivity
              of $\mathrm{Max}_{\mathrm{peak}}$/$\mathrm{Max}_{\mathrm{smooth}}$
              to changes to the different input parameters. The  
              values of the input parameters are the same as shown in the former
              figures in Sect. \ref{measureablecharacteristics}. 
              }}
\end{figure}
Thus the properties of the amplitude spectra we measure with the ratio 
$\mathrm{Max}_{\mathrm{peak}}$/$\mathrm{Max}_{\mathrm{smooth}}$ are mainly 
determined by the damping time. This ratio therefore provides 
a robust measure of the damping time, although it is not very precise 
(due to scatter). 
From Fig. 11 
we see that the most likely damping time, given the observations, is 
$d\simeq2\,$days. 
In Fig. 12 
we show the results of simulations with a set
of input parameters optimized for $d\sim2$--3$\,$days while having the envelope fixed 
at FWHM$_{\mathrm{in}}=81.6\,$\muHz~(cf. Sect. \ref{envelopesensitivity}). 
All eight parameters are 
plotted in a single plot by showing the difference between observations and 
simulations relative to the rms scatter in the simulations. We find the most likely 
amplitude of \Hya~to be $\mathrm{Amp}_{\mathrm{scale}}\sim2.0\,$m/s.

\begin{figure}[t]\centering
\epsffile{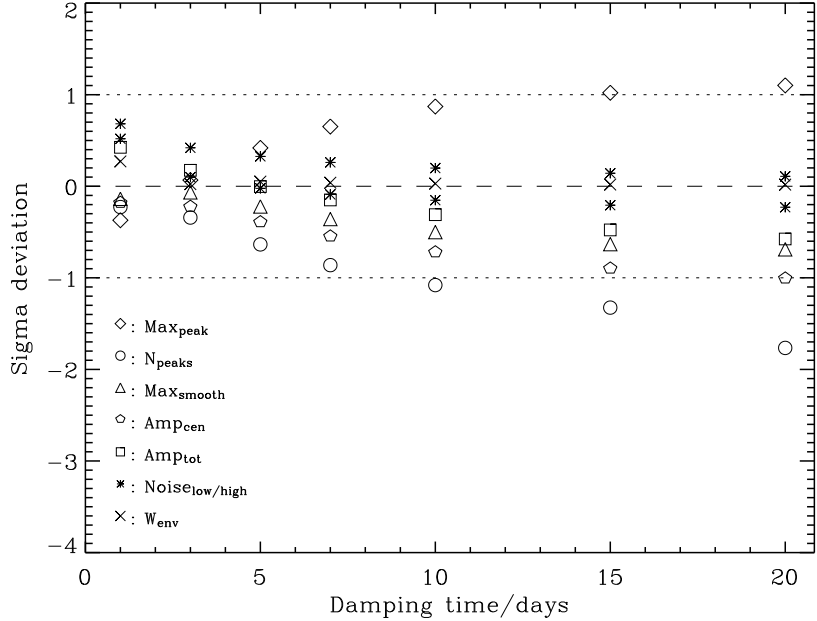}
\caption[]{\label{sigmadevoptimum}
\footnotesize{The deviation between observations and simulations relative to 
              the rms scatter of the simulations (sigma deviation) of all eight 
              measured parameters as a function of input damping time.
              The input parameters (optimized for $d\sim2$--$3\,$days) are: 
              $\mathrm{Amp}_{\mathrm{scale}}=2.0\,$m/s, 
              $x=0.7$ (FWHM$_\mathrm{in}=81.6$\muHz), 
              $\mathrm{Noise}_{\mathrm{white}}=0.10\,$m/s,
              $\mathrm{Noise}_{\mathrm{slope}}=1.5$,
              $\mathrm{Noise}_{\mathrm{scale}}=3.0$, and 18 input frequencies
              with a mean frequency 
              separation of 6.8\muHz.
              }}
\end{figure}

When we optimized the input parameters for a damping time of $\sim15\,$days, we
got roughly the same result as shown in Fig. 12. 
Changing the amplitude or the number of frequencies does not change 
$\mathrm{Max}_{\mathrm{peak}}$ and N$_{\mathrm{peaks}}$ without also 
affecting the other parameters and worsening the fit 
(cf. Figs. 8 and 10). 
The problem with a damping time of $\sim15\,$days
or longer is that we cannot produce enough high peaks while keeping the 
other measures of power down at the observed level. 

It is possible to include a few extra modes (in addition to the 18 modes
with mean frequency separation 6.8\muHz) in the 
simulations and still have an 
acceptable fit with the observations because the change in 
the measured parameters is small, relative to the scatter, provided we also 
reduce the input amplitude slightly 
(see Figs. 8 and 10). 

\section{Discussion and conclusion}\label{discussion}

We have, in the previous sections, described a method for generating 
realistic simulations of stochastically 
excited oscillations, including an arbitrary noise function and an arbitrary 
window function.
Using time series produced by our simulator we showed how the damping 
time, amplitude, and other mode properties of the oscillations could be determined 
by comparing the overall structure of observed and simulated amplitude 
spectra. 
The method was applied to the single-site time 
series of radial velocity measurement of the red giant star 
\Hya~\cite{Stello02,Frandsen02}.

Due to the stochastic nature of the simulated solar-like oscillations, 
we see large variations 
in the amplitude spectra when the length of the time series is not 
significantly ($\gtrsim 10$ times) longer than the damping time. Hence, a large 
scatter is induced for some of the measured parameters that describe the 
characteristics of the amplitude spectra 
(see Fig. 6). 
In the case of \Hya, we are therefore not able
to exclude the damping time of $\sim 15$--$20\,$days calculated by 
\inlinecite{HoudekGough02}. 
However, based on this single dataset, a shorter damping time of only a 
few days seems much more likely (see Fig. 12 
\textit{diamonds} and \textit{circles}).
A clear rejection of a damping time of 15 days or longer 
would require a reduction of
the scatter seen in, e.g., Fig. 11 
by at least a factor 
of two, and hence, using single-site observations, a time series of $\sim150\,$days 
(assuming that the measured parameters from the observations are unchanged).
The optimum fit to the observations, assuming purely radial modes, gave a 
maximum amplitude $\sim2.0\,$m/s, in good agreement with the calculations by 
\inlinecite{HoudekGough02}, and damping time $\sim2$--$3\,$days.
Also, due to scatter, 
it was not possible to exclude the presence of a few extra modes beside the 
18 modes with mean frequency separation 6.8\muHz.
Since the theoretical calculations of the mode inertia 
\cite{Teixeira02privatecommuniation} show that the radial modes should 
be excited to 
larger amplitudes than the higher degree modes, we conclude that our 
simulations strongly suggest that the \Hya~amplitude spectrum is dominated 
by radial modes, but with a possible presence of a few higher order modes.

Below we itemize the limitations of the current 
investigation and discuss their consequences: 

\begin{itemize}
 \item[-] 
          Our approximation of using a damping time that does not vary with
          frequency underestimates the damping time for low 
          frequency modes (by creating more peaks) while overestimating 
          the damping time at the high frequency end (producing fewer peaks) 
          relative to the theoretical damping rate 
          (\inlinecite{HoudekGough02}; Fig.1). 
          Hence, due to these opposing effects,
          this should have little effect on our results.

 \item[-] 
          We included non-radial modes by assuming that they were excited to
          the same amplitudes as the radial modes in the same frequency range.
          A comprehensive treatment of the non-radial modes should include
          both the relative mode inertia, based on a pulsation model, and the spatial
          response of the observations due to projection effects 
          for modes of different degree. The technique presented in this paper
          to measure mode life time is probably not sensitive enough to 
          make worthwhile a more comprehensive treatment of the non-radial modes.

 \item[-] 
          We tested for any significant effect from the 
          adopted deviation of the frequencies from equal spacing by 
          testing both the scatter of 0.2\muHz~seen in the pulsation 
          model (as in the examples shown in 
          Sect. \ref{measureablecharacteristics}) 
          and the larger scatter 
          (0.6\muHz) seen when the observed frequencies were ordered to 
          match the radial modes (see \opencite{Stello02}).
          In these two cases, the eight measured parameters were found to be 
          nearly unchanged, with 
          a deviation of less than $0.1\sigma$, which is the precision with which 
          we know the mean values based on 100 independent simulations.

          Furthermore, we tested a very irregular frequency distribution by using 
          the 13 observed peaks with S/N$\ge 3.5$ (see \opencite{Stello02}) 
          as input frequencies, neglecting possible contamination from alias 
          peaks and noise. The results are similar to those shown in 
          Sect. \ref{measureablecharacteristics}.

          We conclude that the regularity in the input frequencies is not 
          important for obtaining our current result. 
          
 \item[-] 
          The observed velocities of \Hya~were 
          reduced using one reference point
          per night, which produced a high-pass filtering 
          of the time series \cite{Frandsen02}. It would presumably require somewhat 
          different values of the input parameters to match the unfiltered 
          amplitude spectrum. We expect this would mostly affect the 
          input noise and amplitude, by underestimating them, and to a lesser 
          extent the damping time.
   
 \item[-] We chose to use the smoothed observed amplitude spectrum as the 
          frequency envelope because it represents the actual observations.
          We expect that a Gaussian profile could be used for simplicity
          with little effect on the results.
   
 \item[-] We used two successive relatively narrow boxcars to smooth the 
          amplitude spectrum 
          in order to obtain $\mathrm{Max}_{\mathrm{smooth}}$. This method
          preserved the large-scale structure while removing variations 
          on small scales.
          Increasing the boxcar widths by a factor of 2 or 3 did not produce 
          any significant change in the difference between 
          $\mathrm{Max}_{\mathrm{smooth}}$ of simulations and observations. 
   
 \item[-] To test the robustness of our method we applied it to measurements  
          in velocity of $\alpha\,$Cen A \cite{Butler04}. Based on a plot 
          similar 
          to Fig. 11, we obtain a damping time of $\sim0$--$5\,$days. A 
          complete 
          analysis, as shown for \Hya, has not been applied to 
          $\alpha\,$Cen A. 
          Since $\alpha\,$Cen A is very similar to the Sun, one would 
          expect the  
          damping time to be only a few days.  
          Our result is consistent with both the solar value (3--4$\,$days) 
          and 
          the value of 1--2$\,$days for $\alpha\,$Cen A found by 
          \inlinecite{bedding04}. 
\end{itemize}

In general, we see from our simulations that when comparing the estimated 
amplitude (either from scaling or theoretical calculations) with observations, 
the amplitude associated with a star from a single dataset
can vary significantly,
especially for stars with mode life times that are long compared to the length of 
the time series (cf. Fig. 6). 

In a future paper, a more detailed frequency analysis of \Hya~will be presented, 
including an analysis of the scatter of frequencies about a uniform distribution, 
which would test the damping time results given in this paper.

\acknowledgements
This work was supported in part by the
Australian Research Council, the Danish National Research Foundation through 
its establishment of the
Theoretical Astrophysics Center, and by the Fund for Scientific Research of 
Flanders through a post-doctoral fellowship of Joris De Ridder and through 
grant G.0178.02.

\bibliographystyle{klunamed}
\bibliography{bib_complete}

\begin{thebibliography}{}

\bibitem[\protect\citeauthoryear{{Bedding} and
  {Kjeldsen}}{2003}]{BeddingKjeldsen03}
{Bedding}, T.~R. and H. {Kjeldsen}: 2003, `{Solar-like Oscillations}'.
\newblock {\em Publications of the Astronomical Society of Australia} {\bf 20},
  203--212.

\bibitem[\protect\citeauthoryear{{Bedding} et~al.}{2004}]{bedding04}
{Bedding}, T.~R., H. {Kjeldsen}, R.~P. {Butler}, C. {McCarthy}, G.~W. {Marcy},
  S.~J. {O'Toole}, C.~G. {Tinney}, and J.~T. {Wright}: 2004, `Oscillation
  frequencies and mode lifetimes in $\alpha\,$Centauri A'.
\newblock ApJ, submitted.

\bibitem[\protect\citeauthoryear{{Brown} et~al.}{1991}]{Brown91}
{Brown}, T.~M., R.~L. {Gilliland}, R.~W. {Noyes}, and L.~W. {Ramsey}: 1991,
  `{Detection of possible p-mode oscillations on Procyon}'.
\newblock {\em \apj} {\bf 368}, 599--609.

\bibitem[\protect\citeauthoryear{{Butler} et~al.}{2004}]{Butler04}
{Butler}, R.~P., T.~R. {Bedding}, H. {Kjeldsen}, C. {McCarthy}, S.~J.
  {O'Toole}, C.~G. {Tinney}, G.~W. {Marcy}, and J.~T. {Wright}: 2004,
  `Ultra-high-precision velocity measurements of oscillations in alpha Cen A'.
\newblock ApJ lett., in press, astro-ph/0311408.

\bibitem[\protect\citeauthoryear{{Chang} and {Gough}}{1998}]{ChangGough98}
{Chang}, H.-Y. and D.~O. {Gough}: 1998, `{On the Power Distribution of Solar P
  MODES}'.
\newblock {\em Sol. Phys.} {\bf 181}, 251--263.

\bibitem[\protect\citeauthoryear{{De Ridder}}{2002}]{Ridder02}
{De Ridder}, J.: 2002, `Time Series Simulator'.
\newblock Technical report, R{\o}mer Science Data Center, Aarhus Universitet.

\bibitem[\protect\citeauthoryear{{De Ridder} et~al.}{2003}]{Ridder03}
{De Ridder}, J., H. {Kjeldsen}, T. {Arentoft}, and A. {Claret}: 2003, `{The
  EDDINGTON Light Curve Simulator}'.
\newblock In: {\em Stellar Structure and Habitable Planet Finding}.
\newblock in press.

\bibitem[\protect\citeauthoryear{{Frandsen} et~al.}{2002}]{Frandsen02}
{Frandsen}, S., F. {Carrier}, C. {Aerts}, D. {Stello}, T. {Maas}, M. {Burnet},
  H. {Bruntt}, T.~C. {Teixeira}, J.~R. {de Medeiros}, F. {Bouchy}, H.
  {Kjeldsen}, F. {Pijpers}, and J. {Christensen-Dalsgaard}: 2002, `{Detection
  of Solar-like oscillations in the G7 giant star $\xi\,$Hya}'.
\newblock {\em \aap} {\bf 394}, L5--L8.

\bibitem[\protect\citeauthoryear{{Goldreich} and
  {Keeley}}{1977}]{GoldreichKeeley77}
{Goldreich}, P. and D.~A. {Keeley}: 1977, `{Solar seismology. II - The
  stochastic excitation of the solar p-modes by turbulent convection}'.
\newblock {\em \apj} {\bf 212}, 243--251.

\bibitem[\protect\citeauthoryear{{Houdek} et~al.}{1999}]{Houdek99}
{Houdek}, G., N.~J. {Balmforth}, J. {Christensen-Dalsgaard}, and D.~O. {Gough}:
  1999, `{Amplitudes of stochastically excited oscillations in main-sequence
  stars}'.
\newblock {\em \aap} {\bf 351}, 582--596.

\bibitem[\protect\citeauthoryear{{Houdek} and {Gough}}{2002}]{HoudekGough02}
{Houdek}, G. and D.~O. {Gough}: 2002, `{Modelling pulsation amplitudes of
  {$\xi$} Hydrae}'.
\newblock {\em \mnras} {\bf 336}, L65--L69.

\bibitem[\protect\citeauthoryear{{Kjeldsen}}{2003}]{Kjeldsen03}
{Kjeldsen}, H.: 2003, `{Space and Ground Based Data for Asteroseismology}'.
\newblock {\em \apss} {\bf 284}, 1--12.

\bibitem[\protect\citeauthoryear{{Kjeldsen} and
  {Bedding}}{1998}]{KjeldsenBedding98}
{Kjeldsen}, H. and T.~R. {Bedding}: 1998, `{MONS: A proposal for a Danish
  satellite}'.
\newblock In: {\em The First MONS Workshop: Science with a Small Space
  Telescope, held in Aarhus, Denmark, June 29 - 30, 1998, Eds.: H. Kjeldsen,
  T.R. Bedding, Aarhus Universitet, p. 1.} pp. 1--+.

\bibitem[\protect\citeauthoryear{{Kjeldsen} and
  {Frandsen}}{1992}]{KjeldsenFrandsen92}
{Kjeldsen}, H. and S. {Frandsen}: 1992, `{High-precision time-resolved CCD
  photometry}'.
\newblock {\em \pasp} {\bf 104}, 413--434.

\bibitem[\protect\citeauthoryear{{Leifsen} et~al.}{2001}]{Leifsen01}
{Leifsen}, T., B.~N. {Andersen}, and T. {Toutain}: 2001, `{Temporal behaviour
  of radial p-modes}'.
\newblock In: {\em ESA SP-464: SOHO 10/GONG 2000 Workshop: Helio- and
  Asteroseismology at the Dawn of the Millennium}. pp. 63--+.

\bibitem[\protect\citeauthoryear{{Priestley}}{1981}]{Priestley81}
{Priestley}, M.~B.: 1981, {\em {Spectral analysis and time series}}, Vol.~1 of
  {\em Probability and mathematical statistics}.
\newblock London; New York: Academic Press.

\bibitem[\protect\citeauthoryear{{Stello}}{2002}]{Stello02}
{Stello}, D.: 2002, `Detecting solar-like oscillations in the giant star
  $\xi\,$Hydrae. New prospects for asteroseismology throughout the HR-diagram'.
\newblock Master's thesis, Institut for Fysik og Astronomi, Aarhus Universitet.
\newblock Available HTTP:
  http:$//$www.phys.au.dk$/$$\sim$stello$/$Publications$/$thesis.pdf.

\bibitem[\protect\citeauthoryear{{Teixeira}}{2002}]{Teixeira02privatecommuniat%
ion}
{Teixeira}, T.~C.: 2002.
\newblock Private communication; work in progress.

\end{thebibliography}

\end{article}
\end{document}